\documentclass[article]{agujournal2019}
\usepackage{url} 
\usepackage{lineno}
\usepackage{soul}
\usepackage{amsmath}
\newcommand{\nico}[1]{\textcolor{black}{#1}}

\draftfalse

\journalname{Geophysical Research Letters}

\begin{document}

%
%


\title{Mixing and  Geometry in the North Atlantic Meridional Overturning Circulation}

%
%




\authors{Renzo Bruera\affil{1}
, Jezabel Curbelo\affil{1} 
, Guillermo Garc\'ia-S\'anchez\affil{2,3} 
 and Ana M. Mancho\affil{2} 
}



\affiliation{1}{Departament de Matem\`atiques. Universitat Polit\'ecnica de Catalunya.  Avda. Diagonal, 647, 08028 Barcelona. Spain}
\affiliation{2}{Instituto de Ciencias Matem\'aticas. Consejo Superior de Investigaciones Cient\'ificas. C/ Nicol\'as Cabrera 13-15. Campus Cantoblanco, 28049 Madrid. Spain}
\affiliation{3}{Escuela T\'ecnica Superior de Ingenieros de Telecomunicaci\'on, Universidad Polit\'ecnica de Madrid, 28040 Madrid, Spain}




\correspondingauthor{Ana M. Mancho}{a.m.mancho@icmat.es}




\begin{keypoints}
\item Lagrangian Coherent Structures  reveal  the complex mixing geometry  of 3D circulating waters.
\item Supported by this tool we are able to highlight upwelling and downwelling routes across the Atlantic Meridional Overturning Circulation. 
\item This approach seems promising to determine the role of vertical mixing on major ocean currents in the Earth’s climate system.
\end{keypoints}

%
%

%
%


\begin{abstract}
Vertical motions across the ocean are central to processes, like CO$_2$ fixation, heat removal or pollutant transport, which are essential to the Earth's climate. This work explores 3D conveyor routes {associated with} the Atlantic Meridional Overturning Circulation (AMOC). Our findings show the geometry of mixing structures in the upper and deep ocean layers by means of Lagrangian Coherent Structures. This tool identifies among others, zones linked to vertical transport and characterizes vertical transport time scales. We focus the study in two regions. The first one is the Flemish Cap region, a zone of interaction between the major AMOC components, where our analysis identifies a domain of deep waters that ascend very rapidly to the ocean surface. The  second one is the Irminger Sea, where our analysis confirms the existence of a downwelling zone, and reveals a previously unreported upwelling connection between very deep waters and the ocean surface.
\end{abstract}

\section*{Plain Language Summary}
The Atlantic Meridional Overturning Circulation (AMOC) is a complex oceanic convective system  involved in the tridimensional distribution of heat, carbon or nutrients. The horizontal transport across the AMOC is well studied, however,  there are not many  studies exploring  mixing and transport processes across the vertical water column, i.e.,   from the surface to the deep ocean and viceversa. To fully understand this 3D  system, this paper  { follows a methodology that links apparently unrelated  elements such as mixing and geometry. This } approach is very promising to better characterize transport across
the vertical column of major ocean currents and determining
their impact on the global climate system.

%
%

%


%
%
%
%

\section{Introduction}

The  Atlantic Meridional Overturning Circulation is one of the major 
\nico{features} on the Earth's oceans, comprising the circulation of many currents in the Atlantic from the southern hemisphere to the North Atlantic \cite{cessi}.
The  AMOC 
plays a central role in the Atlantic climate.
 For instance, northward ocean heat transport
achieved by the AMOC is believed to be responsible for the relatively warm  winters in Europe \cite{buckmars2016,gulf}.
On the other hand, there exists evidence on how the recent decline of Arctic sea ice weakens the circulation of the AMOC \cite{liu2018}.
{ Vertical transport across the AMOC due to vertical motions plays also an important role in this regard since it is essential for  many oceanic biological and chemical processes.} Indeed, vertical motions act as  primary pathways for transporting heat, freshwater, and carbon from the surface to the deep ocean \cite{kampf, grigory, liang},
and this is also the case for the AMOC \cite{kostov2014,buckmars2016}.

\begin{figure}
\centering
\includegraphics[width=\linewidth]{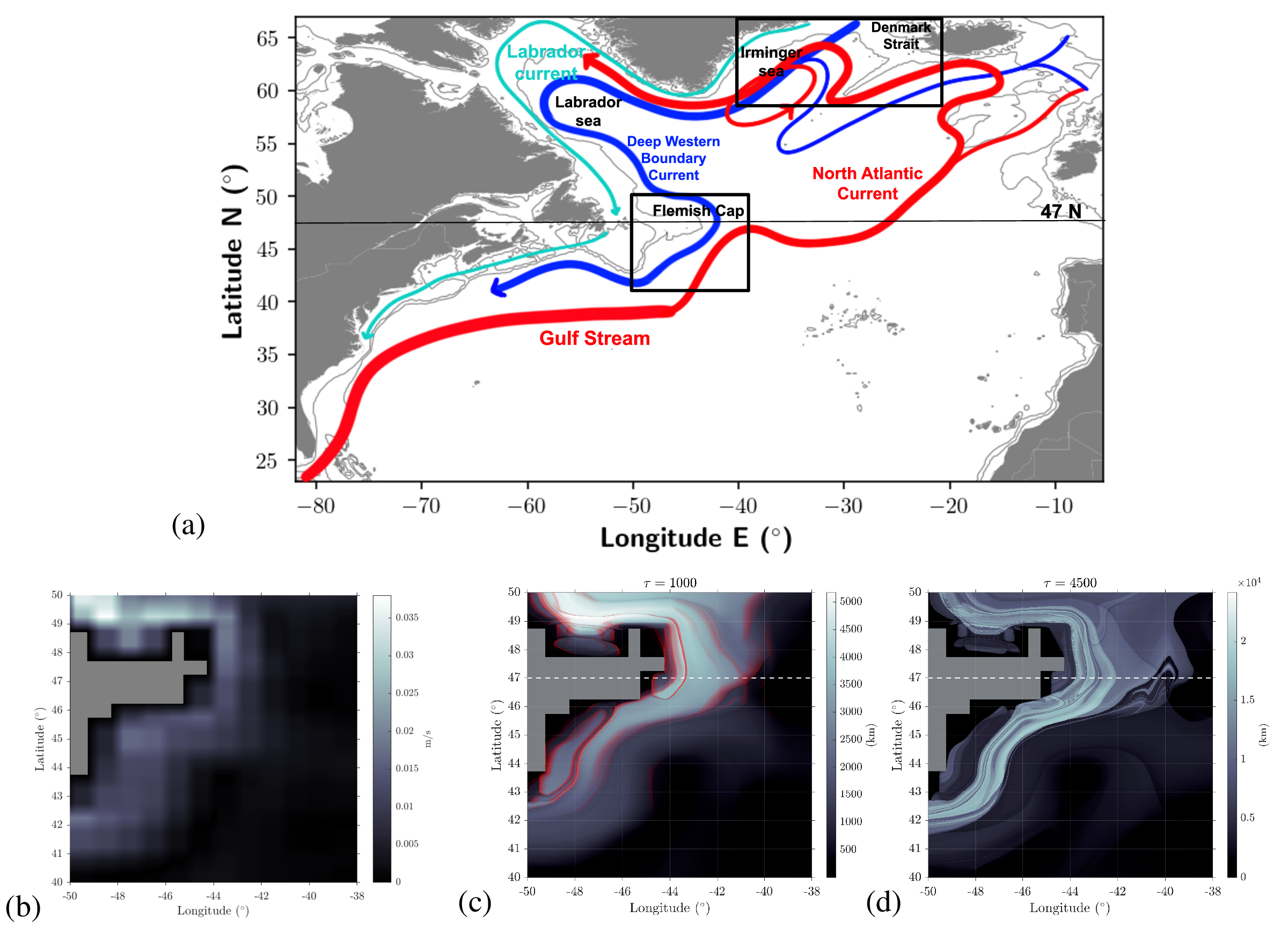}
\caption{{a) Schematic Map of the main component of the AMOC. The black line corresponds to the fixed latitude 47N and the boxes highlight the two regions on which we focus our study: the Flemish Cap and the Irminger Sea and Denmark Strait areas. The major topographical features are included for level curves at 127 and 1200 m depth.  
Bathymetry data was obtained from GEBCO$\_\textrm{2020}$ 15 arcsecond
grid (\url{http://www.gebco.net}). Panels b), c) and d) represent horizontal slices at 1.3 km depth in the Flemish cap region, the first zone of study. b) Modulus of the eastward and northwards velocity components $(u,v)$; (c-d) Lagrangian descriptor $M$  with $\tau=1000$ days (c) and $\tau=4500$ days (d). The dashed line highlights latitude 47$^\circ$N along which a vertical slice is represented in figure \ref{fig:ver}. The gray region corresponds to the bathymetric topography, there is no data for the velocity here.}}
\label{fig:amoc2}
\end{figure}

 {Interior pathways in the AMOC, which  comprises both shallow
and deep currents, are revealed by means of concentrations of observations of chemical tracers \cite{pickart,sme} or  trajectories of subsurface floats \cite{davis,bower2009}.} A sketch on \nico{North Atlantic} transport routes across the AMOC system is presented in Fig \ref{fig:amoc2}a) (see also \citeA{garib}). 
 Red colors indicate warm shallow currents, while blue arrows correspond to cold deep waters. The main element in the shallow layers of the upper cell of the AMOC is the Gulf Stream (GS), which originated in the tropical waters of the Gulf of Mexico and the Caribbean Sea and flows northwards following the eastern coast of the United States. It detaches from the coast at Cape Hatteras and turns northeastward. Near the Grand Banks of Newfoundland, the GS meets the southward-flowing cold waters of the Labrador Current (LC), which joins the GS and together they form the North Atlantic Current (NAC). The NAC continues to flow northeastward while several smaller currents branch off and head southward. On the other side of the Atlantic, the NAC splits into several branches, one of which heads westward, towards the Irminger Sea, and another one keeps flowing to the north through Rockall Trough. {These branches, jointly with  the LC are referred to as the subpolar gyre \cite{little2019}. }

As the NAC flows to the north, buoyancy loss due to cooling {makes} the water parcels sink and form dense, cold water masses known collectively as the North Atlantic Deep Water (NADW). The NADW is usually split into several water bodies, the main ones being the Labrador Sea Water (LSW), formed in the Labrador Sea, and the Overflow Waters, coming from the {Nordic} seas.
Initially, it was believed that the main NADW export pathway was the Deep Western Boundary Current (DWBC), a deep, cold, slow current flowing southward along the southern coast of Greenland and the eastern coast of North America. However, for latitudes below 47$^\circ$N  the majority of the LSW that is exported travels through interior pathways \cite{bower2009}.
{Time motion scales between the upper and mid-depth  AMOC branches are very different}. The GS moves relatively  fast compared to the NADW. Upper waters {travel over} the surface in scales of months, while deep waters circulate in the scale of years.  

Predicting  and describing how transport occurs in the ocean including vertical displacements is  challenging \cite{pascual2020,sca}. Vertical velocities are small, which makes their direct measurement  difficult, especially in the subsurface ocean \cite{liang22}. {In recent years ocean data services    have become available, which among  other variables provide vertical velocities, offering alternatives for addressing these problems \cite{balma}}.
 Studies to understand fully 3D transport across ocean currents \cite{grigory,liang22,ito}, typically have examined averages of the velocity fields over many years, with a special focus on vertical velocities, from which transport pathways have been roughly identified. However this approach is very simplistic because fluid parcels in 3D flows
can follow very complicated trajectories,  even in well-controlled flows such as those in lab experiments \cite{speetjens, wiggins2010}. Indeed,  it has been demonstrated that in these settings Lagrangian transport can be very  intricate and chaotic \cite{psc2010,iri}. 
The present work achieves a step forwards in the understanding of 3D transport across the AMOC by studying the fully nonlinear fluid parcels evolution.  Inspired by Poincar\'e's work, our approach  searches for geometrical structures that  separate regions corresponding to trajectories having qualitatively different dynamical behaviors. 
These geometrical structures,  referred to as  Lagrangian Coherent Structures (LCS) \cite{shadcen2005} act as material barriers that fluid particles cannot cross. In summary, this viewpoint puts into value  techniques coming from dynamical systems theory that provides a simple framework  to understand the  chaotic transport of water masses across the global ocean circulation. These techniques are different from previous efforts in this direction  based on first and last passage time distributions  \cite{transp}, direct statistics on trajectories \cite{cessi2} { or other seeding strategies \cite{fro}}. Our study starts from available velocity products provided by qualified models. In particular, it considers a model for the AMOC circulation realized from 
averaging velocity fields, which are obtained from ECCO (Estimating the Circulation and Climate of the Ocean) products, which is also one of the products used by \citeA{liang22} and \citeA{cessi2}. Our analysis is able to describe in detail the mixing structure of rather coherent water ‘jets’, both  in the upper and deep ocean layers, which eventually mix with surrounding waters in different time scales. We identify previously unidentified upwelling zones and confirm previously reported downwelling zones from the upper to the deep Atlantic. We also identify their mixing time scales. 


\section{Data and Methodology}
\subsection{Data}
This work  analyzes transport {associated with} the AMOC by considering velocity data from the ECCO Version 4 Release 4 (ECCO V4r4) product, freely and publicly available upon registration \cite{ECCO_dataset}. ECCO (Estimating the Circulation and Climate of the Ocean) is a scientific enterprise established in 1999 aiming to produce a quantitative description of the state of the global ocean. It is funded by NASA and it gathers international scientists from a variety of institutions. The ECCO V4r4 contains data about the state of the ocean in the time period ranging from 1992 to 2017. The data is based on the solution of the MIT general circulation model (MITgcm) 
and it incorporates a wide range of satellite and in situ observational data \cite{eccov4r4,eccov4}.

\nico{The data used  is the interp\_monthly version, with resolution  0.5$^\circ$  in both latitude and longitude, and  50 non-equispaced depth levels.
 Each data file contains a one-month average of the velocity field.}
The use of this data is supported by other studies on the AMOC, which  also have used ECCO products.  For instance, \citeA{kostov2021} isolates the sensitivity of the North Atlantic's overturning circulation  to winds, temperature, and salinity and states that ECCO V4r4 reproduces with very high accuracy measurements in the North Atlantic  and also 
the observed subtropical AMOC. Results by  \citeA{liang22} on upwelling and downwelling zones { including the AMOC} are also based  among others on ECCO V4r3 products. 


In order to study the mean transport properties and features of the AMOC, the pointwise average velocity field over the period 1992-2017, has been considered. This period coincides with the one covered by the ECCO V4r4 product. 

\subsection{Methodology}
Fluid parcels in the ocean
follow 3D trajectories ${\bf x}(t)$, which   are solutions to the system: 
\begin{equation}
    \frac{d {\bf x}}{d t}= {\bf v} ({\bf x}, t), \label{eq:DS}
\end{equation} 
where ${\bf v}({\bf x}, t )$ {is the vector field  related to ocean currents, which   are time-averaged, implying that the system} \eqref{eq:DS} is stationary, {\it i.e.}   ${\bf v}({\bf x},t )={\bf v}( {\bf x} )$.
{Equation \eqref{eq:DS} is expressed in cartesian coordinates, where the origin is placed at the Earth's center, and its elements are related to the eastward, northward, and vertical components of the ocean currents. Given that ocean currents are given as data sets, trajectories in \eqref{eq:DS}  are integrated  by  linearly interpolating the vector field  and using a 4th order Runge-Kutta scheme with time step 1 hour (other details at \citeA{curbelo2017}). Beyond the simple trajectory   integration,  we adopt the LCS perspective that provides a partition of the ocean  separating regions in which fluid parcels behave   differently. }
 Of course, the results of our analysis, like those of \citeA{grigory,liang22,ito}, are based on the data set chosen to represent the {velocities}. { Following \citeA{cessi2}, the ECCO V4 product seems particularly suitable for the study of long time scales in the global ocean circulation, and also represents well the transport of major features across the whole ocean and especially the estimate of the Meridional Overturning Circulation (MOC) compared to independent estimates \cite{cessi3}.} 
 
In this work, LCS are computed by means of the Lagrangian descriptor (LD) known as the $M$ function \cite{madrid2009,mendoza2010,mancho2013}, which already   has been used to visualize  three-dimensional Lagrangian structures  in  3D flows \cite{lopesino2017,gg2018,curbelo2017,curbelo2018b,curbelo2018a,curbelo2021,niang2020}.
The function $M$ is given by the expression
\begin{equation}
M(\mathbf{x}_{0},t_0,\tau) = \int_{t_0-\tau}^{t_0+\tau}\|\mathbf{v}(\mathbf{x}(t;\mathbf{x}_0),t)\| \; dt \;,
\label{M} 
\end{equation} 

\noindent
where $\mathbf{v}(\mathbf{x},t)$ is the vector field, $\mathbf{x}(t;\mathbf{x}_0)$ denotes the trajectory going through $\mathbf{x}_0$ at time $t_0$ and $\| \cdot\|$ is the Euclidean norm. $M$ is the arc length of the trajectory traced by a fluid parcel starting at $x_0 = \mathbf{x}(t_0)$ as it evolves forwards and backward in time for a time interval  $(t_0-\tau, t_0+\tau)$. 
 {
The structure of $M$ depends  on  $\tau$.  Figure \ref{fig:amoc2}(b)  represents the  modulus of the eastwards and northwards velocity components in a  plane at a depth of 1.3 km that intersects the upper part of the DWBC. Panels (c-d) represent  $M$  in the same area.
 For small  $\tau$, i.e., $\tau=1000$ days, the pattern is smoother and has less structure (panel (c)) than for larger $\tau$, i.e., $\tau=4500$ days (panel (d)). The emergence of singular features    is highlighted  with the red tone in  (c) }
 
\begin{figure}
\centering
\includegraphics[width=.9\linewidth]{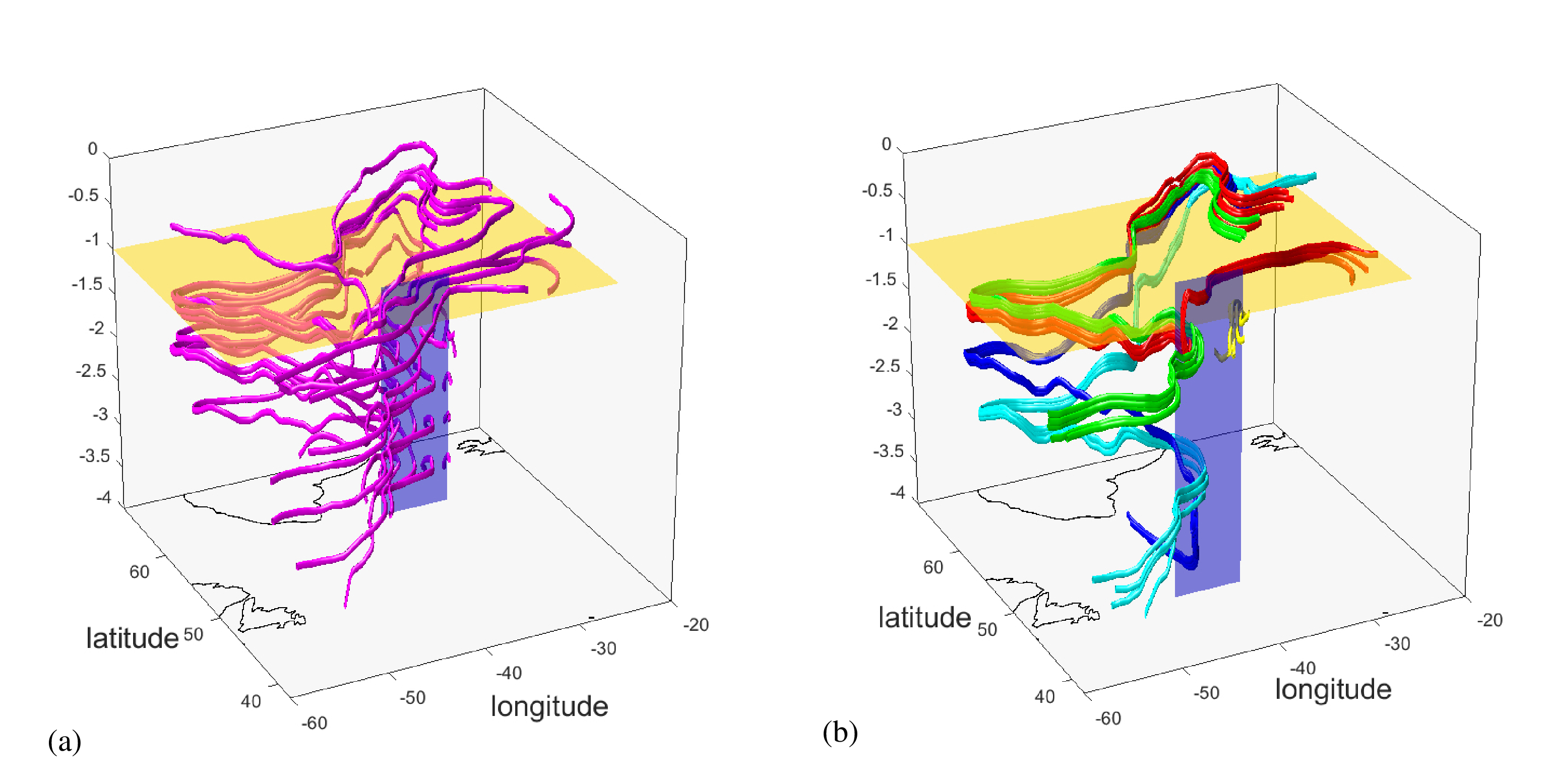}
\caption{ {Forward and backward evolution for a period of $\tau$=2000 days of 25 fluid parcels  placed over the vertical blue plane at 47$^\circ$N.  a) Fluid parcels are located on a regular grid on the plane; b) five clusters of five fluid parcels each, placed in regions identified from the patterns visible from $M$.}}
\label{fig:clus}
\end{figure}

 { These patterns mark boundaries between regions
in which particles have different qualitative behaviors. They are used to select clusters of particles that behave similarly during the time interval.  For instance,  Figure \ref{fig:clus} shows the evolution of 25 fluid parcels whose initial conditions are in the blue vertical plane, located at 47$^\circ$N (white dashed line in \ref{fig:amoc2}(c-d)). In (a) no pattern has been used as a guide to place their initial position, and  are uniformly distributed  on the plane. Their representation  is difficult to interpret with no apparent order. 
In contrast, at (b) colored initial conditions are selected according to the different domains in  $M$ to which they belong. 
Their evolution indicates that {\em particles within the same domain remain grouped} and this is the property  to be exploited by our analysis.  Although the $M$ structure provides a partition where particles behave differently,  a priori it is not possible to know what type of transport is associated with them and this is discovered with  additional integrations per domain.} 

{The function $M$ reflects the transport history
of fluid parcels, and in highly chaotic systems, it is expected to be increasingly complex for longer $\tau$. 
Indeed, within each domain in Fig.\ref{fig:amoc2}(c)  particles evolve in clusters until  time $\tau=1000$. If  $\tau$ increases to 4500,  particles might scatter between $1000<t<4500$, adding new partitions within each partition (see  (d)), but still this would be consistent with their not scattering at $t<1000$. 
The value of $\tau$ must be chosen so that $M$ reveals the relevant transport structures for the problem under study.  Consistent choices for $\tau$ along  this paper are $\tau=2000$ and $4500$ days.}

\section{Results}

For our analysis, we have selected two regions where we have taken sections of the LD supporting the interpretation of transport processes across AMOC. Representations on planes do not imply that the movement of the fluid particles is limited to them, but rather that information about the fluid trajectories is collected in these planes, both forwards and backward in time during $\tau$ periods, no matter what these trajectories are like or which have been the regions visited. The first zone is located at 47$^\circ$N a few kilometers off the coast of Newfoundland, in a region known as Flemish cap (see Figure \ref{fig:amoc2}a) between 45$^\circ$W and 38$^\circ$W. We have chosen this particular location because it is  a zone of interaction between many of the major components of the AMOC: the LC, the NAC, and the GS, and the NADW, whose main pathway is the DWBC. Moreover, it is located within a region known as the “transition zone”, which has been suggested to be central to the understanding of the decadal AMOC variability \cite{buckmars2016}. 
 Therefore, we expect this region to provide some insights into the interaction between the transport pathways in the AMOC and, in future investigations, to analyze possible seasonal and decadal variations in the observed characteristics.

  \begin{figure}
\centering
\includegraphics[width=.9\linewidth]{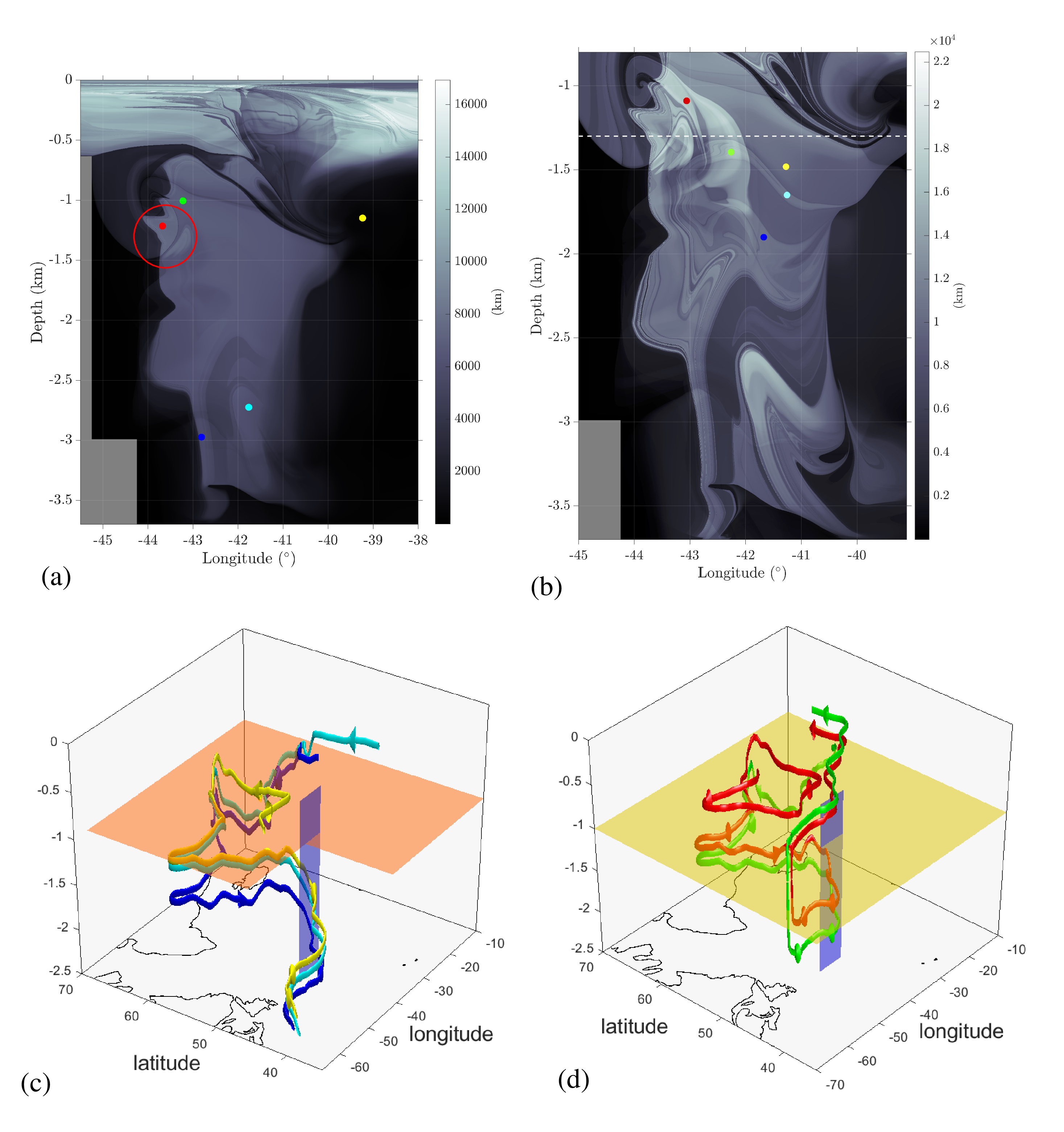}
\caption{Vertical section at constant latitude 47$^\circ$N of  $M$ in the first chosen region, the Flemish Cap  computed with a) $\tau=2000$ days and b) {$\tau=4500$} days. 
A red circle in panel (a) highlights parcel trajectories with ascending behavior described in the text \nico{and colored dots are placed in domains of clustered trajectories of Figure \ref{fig:clus}b)}. Panels (c-d) show the evolution on the Atlantic basin of the fluid parcels trajectories  passing through the selected points marked in (b). Arrows marking the time direction are placed over the trajectories. The vertical blue plane is at 47$^\circ$N and the horizontal ones are at -1km. They are added for visualization purposes.}
\label{fig:ver}
\end{figure}

 Let us start analyzing { this} first chosen region. Figure \ref{fig:ver}(a-b) show the evaluation of $M$ across a vertical section at constant latitude $47^\circ N$ between longitudes $45^\circ$W and  $38^\circ$W and depths expressed in km. This plane contains a transversal section of the DWBC. Panel (a) shows a global view from the top to the bottom for an integration period of $\tau=$2000 days. The image shows an upper layer (in the range 0-0.6 km) with a very rich pattern if compared to the lower ones. On that zone, many  intermingled singular { features are recognized each  corresponding to fluid parcels with different origins or fates, as   expected in a region in which fluid is stirred or mixed. This is in agreement with the ocean mixed layer. }
 The  rather  light gray tone confirms high $M$ values related to large displacements for fluid parcels, which are consistent with faster fluid motions in the zone. In contrast,  deep zones at this integration period present darker features that confirm slower motions \nico{and the need for larger time scales to achieve mixing}. 
 A relatively coherent water ‘jet’ is visible, { between  $44^\circ$W-$40^\circ$W and 1-3 km},  broken by some \nico{structures, where some colored points have been placed. 
 For instance, broken features are identified approximately between  depths  1-1.4 km  and longitudes $44^\circ$W-$43^\circ$W, and are highlighted by a red circle and a red parcel.} Panel (b) displays a zoom into  panel (a)   with an integration period  increased to  $\tau=${4500} days. During this integration period, new structures are identified that further cover  a region that in panel (a) was smooth.  This means that the water stream is no longer a coherent water ‘jet’ during this integration period. When the integration time is long enough, the current  splits into other smaller currents, that correspond to water masses received from other sources or traveling to different fates. 
 Features visible in $M$  serve as a guide to identifying transport pathways across the deep Atlantic current and  its connections to the upper Atlantic currents and are the ones used to cluster fluid parcels in Figure \ref{fig:clus}b). \nico{Indeed, each colored dot in Figure \ref{fig:ver}a) represents a cluster of Figure \ref{fig:clus}. Given that all fluid parcels within each domain behave similarly, we report the results just for one selected trajectory per domain.} Although  the $M$  structure is very rich, \nico{especially at large $\tau$}, we just report here  about transport on picked domains, as an exhaustive description of the behavior on all of them is unmanageable.  We describe those domains associated with the more  interesting transport links between upper and deep Atlantic waters.    
 \nico{Selected domains  appear in } panel (b)  with different colors. Figure \ref{fig:ver}(c-d) display  the evolution on the Atlantic basin of the  fluid  parcels trajectories passing through the selected points highlighted in (b). The  transparent vertical blue plane  marks the position in the basin of sections displayed  in (a) and (b). {A  horizontal plane is placed at 1 km depth separating the upper from deep layers.  Arrows on the trajectories indicate the motion direction. }
 
 A common feature of all fluid parcels marked in colors \nico{in this panel} is that  they have come from the surface, i.e., they have sunk into the deep Atlantic waters coming  from the  upper layer corresponding to waters with depths smaller than 1 km. A remarkable feature of the red and green fluid parcels is that they do not remain on the DWBC. On the contrary, near the North American coast, they rapidly ascend (within 80 days)  to the upper zone, re-circulating into the GS. { This time interval is estimated from the time steps necessary   to integrate the trajectory between the relevant lower and upper  positions.}
 To the best of our knowledge, this ascending behavior has not been previously described in the literature for the AMOC system and appears to  be connected to stir-up effects linked to the bathymetric topography \cite{sci22}, since the positions at which  those trajectories ascend are close to  the edge of the continental shelf { at $\sim$ 64$^\circ$W-42$^\circ$N.  On the other hand,} this evolution is in contrast with the much larger ascending times (above 1000 years),  reported for dissolved inorganic carbon below 1000 m \cite{co2}. 
 \nico{The ascending property of red and green parcels in (b) is shared by the red dot in (a), 
 whose  upwards motion is that of the red cluster in Figure \ref{fig:clus}b). 
 However, the upwelling of the red and green parcels in (b) take longer, appearing only at $\tau=4500$.} 
\nico{ Additional convoluted forms emerge at this point in  panel (b)}, covering both the light feature just described and the smoother darker regions --which expand between  41$^\circ$W-44$^\circ$W and  1-3 km-- completing the mixing pattern between upper and  deep Atlantic waters. These features, contrary to the previous ones that correspond to differences in future time behavior, highlight different backward time behaviors.
 In this case the yellow and red fluid parcels share the backward fate, i.e., both come from a surface circling loop in the mid-Atlantic, at the south of Greenland, while the cyan, green, and blue fluid parcels come from an eastern  branch of the AMOC.
 The integration period used in panel {(b)  provides a time scale of order 12 years (i.e. 12$\sim$4500/365)} for mixing to occur. The horizontal mixing structure of the DWBC at its upper zone (1.3 km depth) is illustrated in Figure \ref{fig:amoc2}. {In Figure \ref{fig:amoc2} (c-d)}, at 47$^\circ$ N between $44^\circ$W-$43^\circ$W, the horizontal section \nico{of features in Figure \ref{fig:ver}b) is recognized.}
  
\begin{figure}
\includegraphics[width=1.05\linewidth]{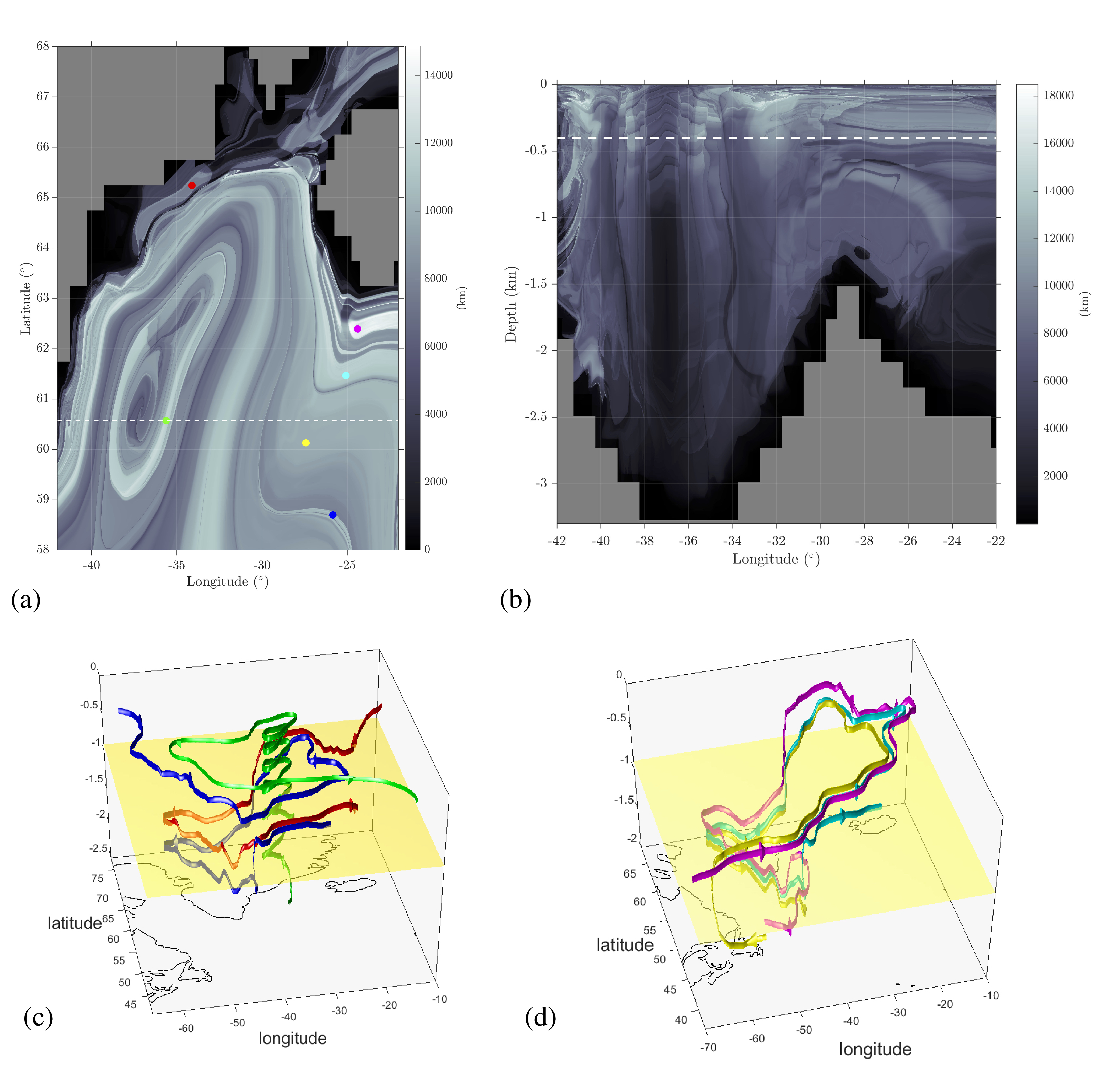}
\caption{(a) {Representation of $M$  in the longitude-latitude plane at 0.4 km depth over the Irminger Sea and Denmark Strait areas with {$\tau$=2000}. The white line represents the latitude ($\sim $60.5$^\circ$ N) at which the computation displayed in (b) is performed; (b)  $M$ in the depth-longitude plane for $\tau$=2000; Panels (c-d) show} forward and backward time evolution in the Atlantic basin of a selection of fluid parcels displayed in panel (a). Arrows marking the time direction are placed over the trajectories. The  yellow plane at -0.9 km is added for visualization purposes. } 
\label{fig:irm}
\end{figure}

{The second region to be analyzed, is the area corresponding to the Irminger Sea and Denmark Strait (see Figure \ref{fig:amoc2}a), where we consider a region between $42^\circ$W-$22^\circ$W and   $58^\circ$N-$68^\circ$N. This is a convective region known for being one of the deep water formation areas for the AMOC. We explore  the vertical motions and
 we spot the sinking of fluid parcels, but also an upwelling region, which to our knowledge has not been reported before.}  Figure\ref{fig:irm}(a) displays the structure of  $M$  in a plane at depth 0.4 km for {$\tau=2000$ days, where  a robust eddy-like structure is recognized. Panel (b) makes visible the internal dynamical skeleton of the eddy-like structure, by showing  $M$  in a vertical plane at  latitude $\sim$60.5$^\circ$ N.   Panels (c-d) display the evolution   of a selection of fluid parcels  visible on frame (a)}. A horizontal transparent yellow plane, placed at 0.9 km depth separates deep and upper Atlantic waters. {Arrows mark the time direction on the trajectories.} Red, blue, yellow, cyan and magenta fluid parcels are associated with water masses that, coming from different AMOC branches at  the surface, have sunk into the deep Atlantic current, {between $\sim$ 41-46$^\circ$W-60-62$^\circ$N. This is located slightly to the north and east of other positions reported \cite{spall}}. The red trajectory comes from the Arctic, passing through the Greenland Sea and through the west of Iceland. The blue trajectory, before joining the GS, circulated through the Labrador Sea. The cyan and magenta trajectories  came  from  the southern branch of the GS. The yellow particle originates in the DWBC and is upwelled near the coast, at around $42^\circ$ N, $62^\circ$ W, where it joins the GS. The behavior of this trajectory is similar to the red and green ones in Figure \ref{fig:ver} (b) and {(d)}. 
  The green trajectory corresponds to ascending counterclockwise circulating  currents coming from depths as deep as 3.5 km that are incorporated into the Labrador current. This ascending motion is slow and takes as long as { 840 days, after which they disperse on the surface. This behavior is similar for all fluid parcels staying in the same partition of the green trajectory. Fluid parcels within the dark feature located at approximately,  37.5$^\circ$ W-60.5$^\circ$ N, also evolve  upward but rotate in movements with a smaller radius.
 The curtain-like structure on the water column  of this eddy, which on panel (b) extends from  $\sim$39$^\circ$-34$^\circ$W, jointly with its horizontal section in panel (a) gives an idea of the volume of this feature that is related to the volume of water transported in this upwelling motion.  } 
  
 
\section{Discussions and Conclusions}

A deep understanding of transport processes across ocean layers is important to gain insights into how heat, carbon, and nutrients are transferred from the surface to the bottom of the ocean and vice versa. Indeed,  these circulations have enormous implications { on  Earth's climate and major  ocean currents, and among them, the AMOC plays an essential role in this regard.} However, vertical motions across these currents are poorly understood, as vertical velocities are very small, and typically the horizontal motion is their most studied feature.    
This study is a step forward  for the better comprehension of fully 3D {stationary} nonlinear transport across the system of currents representing the AMOC. 

We  build a model considering  velocities averages  of available products, such as ECCO,  for long periods of time. Velocities are used to  assemble a fully 3D {stationary} nonlinear model whose solutions represent fluid parcels' trajectories realizing transport across the system.  Transport is interpreted by  LCS, which have provided a mixing pattern  both for the upper Atlantic waters and for the DWBC. As a result of this analysis, we have confirmed that the AMOC possesses downwelling zones placed in the area of the Irminger Sea. We have also identified two upwelling zones, which  have not been previously reported in the literature, of which we   have also characterized their ascending time. The first upwelling  area is placed in the western Atlantic basin, close to the edge of the continental shelf. We find that here deep ocean water rises tied to the structure of the ocean floor, and does it in a relatively short period of time (80 days) in close connections with recent findings reported in \citeA{sci22}.  
\nico{In addition to the previously mentioned effect, there is also a slow upwelling zone in the Irminger Sea that involves water masses taking up to 840 days to ascend due to eddy-induced upward motion. These ascending times are different from those found in a previous study by Primeau (2005), possibly due to variations in data sets or because our study focuses on specific locations of upwelling rather than overall averages across the ocean.}

Our methodology  \nico{achieves a geometrical partition of the ocean and extracts patterns of order from  the apparent disorder of fluid paths, thus helping to  interpret transport through the AMOC. The implementation of this approach in this and other ocean areas supports a sharp understanding of conveyance across the vertical column of major ocean currents. }

\acknowledgments
RB acknowledges support of a CSIC JAE intro fellowship.  AMM and GGS acknowledge the support of a CSIC PIE project Ref. 202250E001 and from grants PID2021-123348OB-I00 funded by 
MCIN/ AEI /10.13039/501100011033/ and by
FEDER A way to make Europe. AMM is an active member
of the CSIC Interdisciplinary Thematic Platform POLARCSIC and acknowledges the support from grant  EIN2020-112235 funded by  
MCIN/ AEI /10.13039/501100011033/ and by the  European Union NextGenerationEU/PRTR. JC also acknowledges the support of the RyC project RYC2018-025169, the Spanish grant PID2020-114043GB-I00 and PID2021-122954NB-I00 and the ``2022 Leonardo Grant for Researchers and Cultural Creators, BBVA Foundation''.

\section*{Data Availability Statement}
The data sets used here are publicly available: ECCO Central Estimate (Version 4 Release 4). They were obtained upon registration from: \\
https://ecco.jpl.nasa.gov/drive/files/Version4/Release4/interp\_113monthly 


%
%



\bibliography{agusample.bib}

%
%
%
%
%

\end{document}